\documentclass[]{spie}

\title{Lessons Learned from Sloan Digital Sky Survey Operations}
\author{S.J. Kleinman\supit{a}, J.E. Gunn\supit{b}, B. Boroski\supit{c},
D. Long\supit{d}, S. Snedden\supit{d}, A. Nitta\supit{a}, J.
Krzesi\'{n}ski\supit{e},
M. Harvanek\supit{f}, E. Neilsen\supit{c}, B. Gillespie\supit{d}, 
J.C. Barentine\supit{g}, A. Uomoto\supit{h}, D. Tucker\supit{c},
D. York\supit{i}, S.
Jester\supit{i}
\skiplinehalf
\supit{a}Gemini Observatory, 670 N. A'ohoku Place, Hilo HI 96720;\\
\supit{b}Dept. of Astrophysical Sciences, Princeton Univ., Princeton NJ
08544;\\
\supit{c}Fermi National Accelerator Laboratory, PO Box 500, Batavia IL
60510;\\
\supit{d}Apache Pt. Observatory, 2001 Apache Pt. Rd., Sunspot NM 88339;\\
\supit{e}Akademia Pedagogiczna w Krakowie, ulica Podchorazych 2, PL-30-084
Kracow Poland;\\
\supit{f}Lowell Observatory, 1400 W. Mars Hill Rd., Flagstaff AZ, 86001;\\
\supit{g}University of Texas, Astronomy Dept., 1 University Station, C1400,
Austin TX 78712;\\
\supit{h}Obs. of the Carnegie Inst. of Washington, 813 Santa
Barbara St., Pasadena CA 91101;\\
\supit{i}Dept. of Astronomy and Astrophysics, Univ. Chicago, 5640 S. Ellis
Ave., Chicago IL 60637;\\
\supit{j}MPIA, K\"{o}nigstuhl 17, D-69117 Heidelberg, Germany\\
}

\begin{document}
\maketitle

\begin{abstract}
Astronomy is changing. Large projects, large collaborations, and large budgets
are becoming the norm.  The Sloan Digital Sky Survey (SDSS) is one example
of this new astronomy, and in operating the original survey, we put in
place and learned many valuable operating principles.
Scientists sometimes have the tendency to invent
everything themselves but when budgets are large, deadlines are many, and
both are tight, learning from others and applying it appropriately can make
the difference between success and failure.  We offer here our experiences
well as our
thoughts, opinions, and beliefs on what we learned in operating the SDSS.

\end{abstract}

\keywords{SDSS, Observatories, Operations, Lessons}

\section{Introduction}
The Sloan Digital Sky Survey \cite{yor00} was an ambitious and 
largely successful effort to conduct a homogeneous survey
of a large piece of the sky --- some seven to ten thousand square degrees in 
five
photometric bandpasses, supplemented by extensive optical spectroscopy of
$\approx 10^6$ galaxies, $10^5$ quasars, and $10^5$ stellar objects.
It was a very coordinated, unified survey involving precise
photometry and follow-up spectroscopy on a specifically chosen region of
sky, mostly in the northern Galactic cap. 
Although not meeting its entire set of initial
goals during the official five year baseline, the shortcomings were few,
the ultimate successes many, and we learned from both.   The responsibility for
the success lies in many,
many talented people who contributed significant parts of their time, energy,
and careers in the effort.  In addition, the project benefited from an
excellent purpose-built telescope \cite{gun06} and matching set of unique, 
state of the art instruments:  a large-field imaging camera \cite{gun98},
dual multi-object fiber-fed spectrographs \cite{sto02}, and an automated
photometry telescope with near real-time data capability, along with extensive 
photometric \cite{fuk96,hog01,lup01,smi02,tuc06}, astrometric \cite{pie03}, 
and spectroscopic systems and pipelines.  The latest data release paper,
that for Data Release 6, is Adelman et al.\ (2008)\cite{ade08}.

In the hopes that some of this work will be of use beyond its immediate
benefit to the SDSS, we attempt here to describe SDSS operations, paying
particular attention to the site observing and engineering efforts and the
many lessons we learned during the survey. 
We endeavored to quantitatively measure operations
efficiency during the survey and used these metrics to help determine our 
operational strategies. Boroski et al.\ (2002)\cite{bor02} discuss the development of the SDSS
operations baseline in more detail as well as
the importance of systems engineering and external reviews in early SDSS
commissioning and operations. 
Not everything in operations can be
strictly quantified, however, and we therefore had sometimes to rely on 
our accumulated experiences, wisdom, and philosophies as well, some of
which are promulgated here.

\section{Goals and Specifications}
One of the SDSS's fundamental strengths was the careful planning 
and specifying of the survey from the early days of the
project\footnote{http://www.astro.princeton.edu/PBOOK/}. The
SDSS's mission, goals, and requirements were carefully detailed and
constructed so that the ultimate survey was an intricate weaving of
details, carefully orchestrated to achieve the desired goals. The survey
telescope, instruments, and sky coverage, for example, were jointly
optimized to efficiently achieve the desired scientific results in terms of both
money and time.

This careful specification process gave operations a very strong base from
which to proceed.  The project's specifications were translated to
form quantitative measurements to evaluate the survey's state and progress.
This pre-defined survey structure allowed us to evaluate proposed changes
against stated requirements, keeping the mission focused and on track.
While the temptation to continually expand the
scientific capacity of our data was large, we did not want to end up with a
dataset that became {\it n} smaller datasets, each a homogeneous subset of an
overall, inhomogeneous dataset.  Thus, the deciding principle was that in 
general, such
improvements were made only if they had little or no impact on the original
data quality objectives (positively or negatively) and did not affect
efficiency too negatively.

Within the SDSS imaging survey, the initially-formulated data quality
objectives, as well as the camera and software realizations, were complete
enough to prevent many extraneous new capabilities and improvements being
tacked on to the survey.   The exceptions to this clarity were within the
calibration and reduction processes, both of which could be applied to the
entire dataset once perfected, thereby assuring continued 
uniformity.  At least in theory.  As it turns out, spectroscopic 
targeting was affected by changing 
photometric reductions.  An object assigned a fiber in one version of the
reductions, might not have been in a later version, and vice versa.
Thus, there are non-uniformities of varying degrees in the
spectroscopic sample selection due to these changes.  In addition, if any
part of the data is eventually found lacking during
final reduction (like additional calibrations), it may be too late to go back 
and acquire the necessary
data or information to fix it.  Thus, even data reductions have to be at a 
fairly complete state early on to fully assess the quality and
completeness of the incoming data.

Although well defined in the original survey specifications, the SDSS
spectroscopic operational objectives 
developed much less lucidly than did
those for the imaging survey.  The main ``problem''
was that the signal to noise of the gathered spectra
was far in excess of what was needed to satisfy the survey's main objective: 
red-shift determination.   We were also late to develop testable functional
requirements for the immense and varied science beyond this objective.
The net result was that the ultimate capability of the
spectroscopic survey was determined primarily by efficiency --- how much
could we do and still finish the survey in a reasonable time.  Much
effort and many misunderstandings could have been avoided had the real
value of the spectroscopic survey beyond mere redshifts been initially
included and fully specified at the start of the survey.  

As with the imaging,
we also went through several attempts at determining the best way to
calibrate each plate's spectra during the night.  We ended up taking a
significant amount of calibration data that ultimately never got used.

Lesson: Have well-specified project objectives which include
a scientific motivation, quantitative objective, and a proposed test at
the start of the project.  Evaluate all activity and proposed changes
against these objectives.  Fervently avoid mission creep.  Plan carefully
for data calibration.

\section{Baselines and Time-Tracking}

A set of goals and objectives become meaningless if they are not
continuously measured and held up for review.  This effort of establishing
baselines and tracking performance
is especially relevant for a time-limited (five-year) project
like the SDSS.  To this end, we made initial time estimates for
all observing operations, including overhead due to instrument change, 
telescope slewing,
centering, focusing, etc.  With these estimates, we constructed a
survey completion baseline and regularly evaluated this against current
progress.

Once survey operations officially started, we 
soon found that we were not meeting all of our benchmarks.  There were some
obvious errors in our baselines, 
but other causes were less easy to identify
and needed a more detailed breakdown of both the baseline and the used
operational time to track them down.
We tried
various ways to gather the latter data, but the only one that really worked
was to 
develop automatic, accurate time/event-tracking
directly into our
operational software\cite{nei02}.  Every primary
command automatically recorded its start and stop times in a running log
file.  We developed additional software to extract these data and
compare them to the detailed baseline. With this information in hand, we
could then
direct our resources towards specific problem areas and
rapidly move towards meeting or exceeding our incremental baselines. The
data also allowed us to more precisely test our baseline assumptions on
nightly weather conditions.  The results from regularly
monitoring baseline progress, including our improvements in problem areas, 
allowed us to demonstrate to ourselves and just as importantly, our 
sponsors, continual
progress and improvement. 

The spectroscopic survey benefited enormously from this attention to the
baseline. Once we had a clear understanding of both the baseline and our data,
we realized a fundamental mistake: we had been observing to the baseline
exposure time estimate, not required signal to noise ratio. In good
conditions, we could achieve the desired signal to noise in less time than
was in the baseline. In worse conditions, of course, we needed more time. Once
corrected, this
realization helped us immensely in understanding our baseline and in providing
a quality, uniform spectroscopic data set. Boroski et al.\ (2002) further 
discuss the
development of the SDSS
operations baselines.

Lesson: Establish an expected baseline for all operations. Build automatic
task time-tracking into your operations software along with corresponding
time-analysis code from the beginning of your project.  Continually
compare your real efforts to your baseline,
but be sure you are comparing against and
measuring the right things.

\section{Systems and Data Telemetry}
Just as important as quantifying our operational efficiency was our effort
to establish baseline operating parameters for our hardware as well.
Many different institutions
and people developed and built the SDSS hardware, yet
it all had to be
maintained, troubleshot, etc. by local personnel on the engineering and
nighttime operations staffs.  Establishing how things look when
behaving properly, therefore, was just as important as having the ability
to see how they are when misbehaving.

We created a very detailed system of data logging using EPICS
(Experimental Physics and Industrial Control 
System\footnote{http://www.aps.anl.gov/epics/}) we called the Telescope
Performance Monitor, TPM \cite{mcg02}. This system
monitored many different signals including telescope positions, drive
currents, temperatures, etc.  It was capable of frequent sampling rates and
realtime display of the data as well as archival plots and trend analysis.
Time and time again, the TPM proved invaluable in solving problems at the
telescope.

During early equipment troubleshooting, we learned three things about
our initial implementation of the TPM:
1) not all necessary signals were being (properly) recorded at the appropriate
frequencies, 
2) the data were not readily accessible for quick analysis and correlation
with other
data, and
3) it was not always clear which signal was what; that is a signal's name
did not always unambiguously identify its nature.
During further system development, therefore, we systematically 
tested and reviewed each measured signal and created a 
generic set of data analysis tools to allow the most
commonly-performed analysis to proceed quickly.  We started documenting the
signal names and what each was measuring, but never fully-developed this as much
as we could. In the end, therefore, the TPM was of most use only to engineers 
familiar
with a given system's specific details.

While this EPICS-based system monitored low-level signals, mostly 
from the telescope and infrastructure, we created another system 
to monitor instrument and data quality status.  The SDSS instruments
all had built-in communication channels to relay instrument status and operating
conditions.   We created master server programs which controlled
communication to each instrument and relayed the status bits to another
program called the Watcher. The Watcher also talked to the data acquisition
computers and did some data analysis of its own to be able to report on the
current status of instrument well-being and data quality. When a parameter
became out of specification,
it sounded an alarm, notifying the appropriate
people of the problem.  One problem in our implementation of the alarm
function was that it only notified people via email and a red button on its
software display panel. If no one was reading email or running the display
software, or worse, the network was down, alarms would get missed. A better
implementation would include other contact methods such as telephone,
pagers, etc.  

Since an alarm that is always alarming really ends up alarming no
one, alarm thresholds were carefully adjusted 
to meaningful levels during the course of the project.
Equally problematic were
alarms that indicated problems with
no immediate solutions nor required actions.  While many
such conditions may indeed need to
be monitored and addressed, they do not need to be handled the same
way as alarms that require immediate action.  We never quite fully
addressed this problem, but our thought was if there is no action that can
or needs to
be taken now, don't alarm.  Instead, find some other way to record 
the problem so that it can be analyzed at some convenient time in the (near) 
future.

Other ways that alarms got ignored included there being no one clearly
responsible for handling the alarm, or no established
procedures for how to address the alarm. We therefore tried to
establish procedures which described who is responsible for handling the
alarms, how to indicate an alarm is being addressed, and finally, what to
do to fix the
cause of the alarm. 

At first, the Watcher merely displayed alarms and informed people of
problems. This service was very valuable, but we soon realized we needed
more.  For example, if the telescope stopped moving during a slew or track,
the watcher would clearly indicate that the telescope had stopped moving,
but not always why it had stopped. Because the SDSS telescope is
highly-interlocked to protect it, its instruments, and its users, there
were many root causes that might cause the telescope to stop moving.  
Since the Watcher already had access to all the
low-level interlock signals, we added the interlock logic within it,
thereby making it possible for the Watcher to 
indicate the root cause of all interlock-driven problems as well as the
problems themselves.  Thus, the
Watcher became not only a problem notification tool, but also a problem
diagnosis tool.  It is unclear which mode was most valuable.  Both were needed
and relied upon extensively.

%A second problem in our implementation was a reliance on the aforementioned
%instrument communication servers. If these died or hung-up, the Watcher
%would lose its ability to monitor things.  We did alarm on loss of server,
%but still, sometimes network problems that stopped communications to the
%server, also stopped the Watcher from communicating the error.

%The existence of these two separate systems, the TPM and the Watcher,
%each based on different dependencies and architectures was either an
%un-necessary complication, or a nice way to limit exposure, depending
%on your point of view. It probably would have helped had the two systems
%been more united and less vulnerable to power and network outages.
%
Lesson: Hardware telemetry is extremely valuable, but make sure
the data are accurate, sampled properly, and easy to analyze.  Don't wait
until there is a problem with the equipment before you test for problems in
your telemetry.
%Monitor as many signals as you can automatically for trouble. 
Alarm when values get out of specification, but be certain the alarms are heard
and not being sent when urgent action is not needed.  Ensure alarms have
known remedies and response procedures.  In complicated systems, computer
diagnostics can greatly help debugging and can often pinpoint the low-level root
cause of a much higher level symptom.  Document what you measure.

\section{Personnel and the Scientist Dilemma}
Astronomy projects often need very specifically-skilled people to play
largely support roles.  
Scientists are not always needed in all skilled positions, 
but when they are, they present an additional complication: they usually
want to do science.
Rewards and professional development need to
be included in their work plan.  The conflicting needs of the project
(support) and the desires of the scientists (science) form what
may be referred to as the {\it scientist dilemma}. 
The scientist dilemma occurs whenever highly-skilled, scientifically
motivated people are needed for support work. This work could be, as in the
case discussed here, operations and observations, but the dilemma
applies equally well to programmers, data analysts, archivists, etc.  

\subsection{The Dilemma}
The SDSS collaboration realized early on that Ph.D.-level people were going
to be required for nightly operations.\footnote{It wasn't so much the
degree itself that was necessary, but several factors that come with it:
observing experience, data handling and analysis, scientific context,
problem solving, and an exposure to scientific computing environments. It
is certainly possible to find these skills and experiences in someone
without a Ph.D., but they are more common in those with it.}
The telescope, instrument,
software, and data systems were complex enough that a high level of skill
was demanded to successfully use and develop them.  In addition, the
nightly observing plan was flexible enough to the current
conditions that scientific tradeoffs between different courses of action
would need to be evaluated in real time to optimize each night's
observations.
We also realized that a stable group of skilled observers
could not only hone the operating systems and procedures to improve 
both efficiency and data uniformity,
but could also
take over some of the software and hardware development work as well, more
finely tuning the initial efforts to fit real observing 
conditions.  This work resulted in continual operational
efficiency improvements and left the system in such a state that by
the end of the project, Ph.D.-level scientists were no longer required to make
operations successful.  This result also helped address the
endgame issue mentioned in the next section.

The problem with this approach is that whereas the project
wanted Ph.D.-level astronomers to learn and understand the complex
operational systems, spend non-observing time improving the systems and
performing required auxiliary tasks (instrument calibration, data
integrity checks, etc.), decide coherent efficient nightly observing
strategies, and operate the telescopes and instruments nightly during
observations,  most Ph.D-level astronomers want to do (at least some)
astronomy --- hence the dilemma.

The only way to really address this dilemma is to simply staff accordingly,
allowing your professional staff
enough time to do their three main tasks: in this case,
observing, system verification and development, and scientific research.
Without the latter, not only do you not have happy workers willing to
devote themselves to the project for its duration, but you also leave them
with no career path beyond future, non-scientific support work. 
%As have others, we also found that those with a vested interest
%in a project usually see the project through with better quality support
%than those without.  Science time and career growth can provide this interest. 

The inevitable argument against this obvious solution to this critical
dilemma is money.  Hiring more people costs more money. Since projects are usually
run with less money than they really need, additional personnel are often
deemed an un-obtainable extravagance. There are several arguments against
this position, though:
\begin{itemize}
\item{When you factor in 
project downtime, intervening overtime, recruitment, relocation, 
training, etc., replacing an employee is expensive.  It can be
cheaper to hire an additional employee and keep the ones you have than to
constantly have to bring-in new replacements as current employees move on
to protect their careers.}
\item{Having skilled observers directly participate in developing 
operations not only increased project efficiency (thereby
decreasing operations costs), but provided staffing in an area that
otherwise would need to be staffed additionally. }
\item{And finally, the personnel costs we are discussing here are not really an
extravagance. They simply represent the real cost of employing scientists.
Scientists come with additional overhead which budgets need to reflect.} 
\end{itemize}

Ultimately, the observers were staffed in such a way that roughly half their
time could be spent observing, a quarter on systems verification and
development,
and a quarter on professional development and scientific research. 
We also created a rotating month away from
assigned observatory duties (roughly annually), so
dedicated time could be spent either starting a new project or pushing an 
ongoing one 
out the door to completion.  Correspondingly, we made available a small amount 
of financial resources for these activities.

The endgame is also an important concern for project staffing. Obviously,
it is advantageous to the project to ensure their key people stay until
the very end of the project.  However, given the time it can take to find
jobs in astronomy, people simply will not be able to stay to the project's
very last
day if there is no guarantee of some additional employment
past it. 
Staffing plans must therefore  be made to provide useful
employment for key employees for some time past the project's end, or
otherwise provide for bringing in, or doing without, last minute replacements.  
Loyalty alone cannot be
relied on.

Lesson: Hiring scientists comes at a cost: they must have some time and
resources to do science.  
Replacing staff is a time-consuming, costly affair. It 
can be cheaper to keep good, happy, staff, than make do with
a rotation of less expensive short-timers.  Keeping people to the end
of a mission necessitates ways of guaranteeing employment past said end.

\subsection{The Management Corollary}
Scientists are not usually trained as mangers and managers are not usually
trained as scientists.
There are some talented people who can play both roles, but rather than
relying on the exception, it is safer to plan for the more commonplace
scenario. 

Like software engineers, professional managers exist for a reason. They
are trained in evaluating personnel, logistics, scheduling, fundraising,
etc.  --- all things not usually found on the transcript of your average
scientist. On the other hand, they are not always well-versed on the science
of their missions and less able to make well-informed compromises between a
project's logistical and scientific needs.  The Large Synoptic Survey Telescope
project\footnote{http://www.lsst.org} is addressing this problem by putting both a
trained scientist and an experienced manager in each management box of their
organizational chart\footnote{http://www.lsst.org/About/lsst\_team.shtml}.
This approach seems sound and time will tell how well this works, but the
important point is to recognize that science leadership and management
leadership are two different things and it is rare to find someone
sufficiently
effective in both.

Similarly, scientists and not usually engineers, and engineers are not
usually scientists. This same discussion applies to a variety
of possible job titles. Hire people to do what they do best, but help them
work together by appropriate levels of management, communication, and
training.  Management and technical training for those that don't have it
can be quite beneficial.

Finally, in multi-institution projects, rivalries, jealousies, and competitions
often arise amongst the different institutions.  These things can sometimes
actually be beneficial to a project, but
at other times, they must be fairly and professionally
curtailed before the project derails itself from within.  
Project-level
management, therefore, needs to strive to be above institutional
partisanship. While it is natural and often un-avoidable to have project 
management come directly from an individual member institution, it is
important for management to work hard to not appear (and more importantly,
to not {\it be}) too attached to an individual partner institution.

Lesson: Recognize the strengths and weaknesses of your management and
work to temper both. Both technical and logistical leadership are needed
throughout the project. Hire the skills you need.  Avoid management with strong institutional biases in
a multi-institutional project.

\subsection{Staffing and Scheduling}
Astronomy operations necessitate shift work and scheduling shift work can
be difficult.
In addition, despite common procedures and training, different observers will
inevitably collect the data in different ways, possibly affecting
data uniformity and efficiency. Since the SDSS observers worked in pairs,
we decided to create a schedule where
roughly half the observing nights of each observer were spent with the same
partner, and half with the others. This arrangement allowed for each team to develop
some cohesiveness and efficiency together as well as to propagate new ideas and
procedures to the rest of the team and integrate out any major differences.  

In addition to the standard night shifts, we also established an
afternoon-evening swing shift
position. This position was rotated into, each observer taking a turn
in it, and offered a few salient features: 1) it allowed telescope and
instrument setup to begin early in the day so any problems could be
addressed before observations, 2) it allowed the night staff to come in
later than would have otherwise been practical had they needed to do the setup
themselves, helping to make the long observing nights more
tolerable, 3) it allowed an overlap between the 
day staff and the observing team to discuss daytime work and other
observing-related issues, and almost as important, 4) it allowed the 
night-working observers an
occasional chance to work mostly in daylight.

We realized early on we needed another unique position among the observing
staff: a coordinator of some sort.  This person could serve as a conduit of
information between the observers and the rest of the collaboration,
bridging the day/night gaps and providing representation of the
observing staff to the rest of the project and the rest of the project to
the observing staff.  This capability was vital in melding the
project's needs with on-site operations.  With emails and project
meetings accumulating throughout the day, it could be difficult for the
night observers to discover what new information they needed to know from
the day's activity for their upcoming night shift. This coordinator
therefore filtered all the day's information and created a way for the
night's observers to ``pull'' the information they needed at the start of
each shift.  This person also worked some night shifts, but only half as many as
the rest of the team, to spend remaining hours in daytime work, accessible
to the rest of the collaboration and at least the first half of the night
time shifts.\footnote{It is probably worth noting that the lead author
of this paper held this coordinator position during the SDSS-I survey and
came to view it as vital to the success of the project, greatly helping
with project communications and establishing uniform operations procedures.
Thus, many of the lessons stated here come from the viewpoint of this
position, although additional experiences at other organizations and discussions
and input from other project members contribute as well.}

Lesson:  Schedule operations staff so they can establish rapport and
efficiency with each other, but allow for cross-pollination of new ideas
and procedures.  Create some part-time daytime positions amongst
the night staff to provide vital overlap and communication with the local
day crew and the project at large.  A single coordinator, or ``lead
observer''' served us well as a single voice for and conduit
between the observing staff and the rest of the collaboration.

\section{Product Delivery and Handover}
While the SDSS may have employed highly-skilled observers and telescope
engineers on site, most of the project's
developers were off-site. Thus, when a new system was delivered to the
site, it had to be taken care of by the site staff, and usually not by the
system developers.  Each delivered system should therefore have had a handover
document establishing the required deliverables (documentation, spares,
procedures, training, etc.) before the developer was allowed to
disavow responsibility for it.  As it turns out, though, the
SDSS never formally established this approach and 
we therefore often relied on developers for continued system support,
sometimes well
after they were on to other projects.
Also, local staff often spent more time than might otherwise have
been necessary when working with these systems.
In rare cases, we simply went without desired changes to a
system because we could find no way to implement them ourselves.

Lesson: Establish formal handover procedures for all
deliverable systems.  Strive to take complete ownership (it may not always
be possible) of a system on-site once delivered.

\section{Software and Version Control}
We start this section first with the lesson.

Lesson: Software is difficult.  It takes time.  It often takes trained
professionals to get done right.  It requires user input and evaluation
throughout development. Plan for an active software life after delivery. 

\subsection{Development}
When computers were expensive and not very powerful,  most things were
done in hardware, or not at all.  These days, though, with incredible
amounts of power available on our laps and desks, software does so
much more than it used to, but too often, only old-style resources and
requirements are allocated to the effort.  Software needs requirements,
management, and a testing plan, just like hardware.  Ideally, it should test the limits
of the hardware from the beginning and be integrated into hardware and
other software acceptance testing and baseline measurements.
%The ``traditional''
%model of hiring a post-doc or two to write all the software rarely works in
%today's larger projects.

We solve problems in software because it is usually
more flexible and quicker to develop than are hardware solutions.  These
advantages, though, often lead to the temptation to purposefully not
define a full set of requirements from the start, relying on software's nature 
as a fast,
flexible solution to provide a last minute product appropriate to
its task.  While there are controlled versions of this approach often
referred to as {\it extreme programming}\footnote{See, for example,
http://www.extremeprogramming.org.} which can indeed
be beneficial, allowing software requirements to change with time
as a project progresses nearly always guarantees that it will be the
last thing done.  If you mean to take an extreme
programming or agile approach to software development, by all means, do so. It
can be a quite useful powerful approach.  But if you are simply allowing
the rest of the project to eventually define your software in order to
save time at the start by not fully specifying a set of requirements,
be aware that this means your software will probably be over budget and
behind schedule, practically by definition.

Many astronomers have significant experience with data reduction software;
fewer, with operations software.  The two are fundamentally different,
however, and skills at one do not necessarily translate to the other.  The
SDSS was fortunate to have some extremely talented programmers, but 
few were really experienced in operations software and few would
call themselves professional programmers.
As such, they often had other priorities besides code ease of use and 
maintenance, efficiency, and general ability to function properly in a complex, 
interactive, time-critical
environment.  In addition, while they did a reasonable job of
predicting the future model of operations and its
environment, less thought was given to how to correct the software once
parts of the
model were inevitably
proven wrong by real operations. 
Plans were made for
software delivery, but not for its continued maintenance and development.
Through a
devoted effort of some original developers, newly-allocated resources, and
contributions by
the observers, the operations software did become reasonably
complete and efficient,
but the path there was often painful.  The salary of an experienced, on-site
programmer prepared to handle the real-word changes/improvements necessary
in the ``delivered'' code, would probably have been paid for several times
over by increased productivity and efficiency in operations throughout the
project.

\subsection{Version Control}
Proper source and binary version control is essential for smooth operations
and problem tracking, both realtime and post-mortem.  Some SDSS developers
used version control systems religiously, some sometimes, and others, not
at all.  Initially.  Software management and problem tracking/solving took
a huge step forward once we initiated a mandatory version control
policy.  All software version numbers were reported in the appropriate
night logs and FITS headers automatically, making the identification of
problem data caused by software bugs much more straightforward than it
would have been otherwise.

We used the widely-available {\it cvs} package for source code version
control, and the less-widely known, but equally valuable FermiLab product,
{\it ups/upd}\ \footnote{http://www.fnal.gov/docs/products/ups/}, for binary 
control.  Both packages allowed us to always know
what version of the software was running when, to quickly revert versions
when necessary, and to record what changes were made to each version.
These practices were absolutely necessary for effective software management
and data quality assurance.  

The version control creed became so deep in the SDSS culture that even our
procedures and documentation were version controlled. In this way, we could
trace changes to the procedures and correlate them with data taken
on any given night as well as allow all the observers
the freedom to edit and
improve them, all in a documented, controllable manner.

\subsection{A Time for Testing}
The observers ultimately controlled the procedures/documentation repository
as well as the binaries that were being run at any given time.  They tested
new builds of code only on non-survey, bright time engineering nights.  The developers
proposed tests for their new changes and the observers tried to augment
those with their own regression tests (although this system was never as
fully-developed as it should/could have been).  Developers tried to be
available during testing, should there be a fix needed.

A new version of code was only
declared ``current'' and used during the science run
when it passed all its tests during the scheduled engineering
time. If it did not pass, the previous software version remained current 
and the developer could work on fixes for the next engineering time.
Because the survey did not operate during the peak of the monthly full
moon time, we created two opportunities for software testing, one at
the end of each monthly run before the bright time and one at the beginning
of the next, just after.

These engineering nights might take away from potential science time in
other projects, but they could still be worth the loss. The set schedule
allowed for clear start and stop dates for operational and hardware
changes and provided opportunities to regularly test the observing systems
for potential problems before they became too large.

Since our engineering night tests could sometimes leave bugs un-discovered
in the new software, we relied on our binary version control system
discussed earlier to allow the flexibility to revert to an older version
during the night if a fatal bug was found.  
Only under rare circumstances (critical bugs that could not be reverted out
of and negatively
impacted data quality or operating efficiency) were new versions of code
tested outside of engineering time.
Bugs that were discovered
during science nights, but that could be worked around, were.  They were then
(hopefully) addressed the following engineering cycle.

Lesson: Software is hard. Budget time and resources adequately.  Allow
and plan
for end-user involvement in the specifications as well as post-delivery
support and further development.
Integrate software version control project-wide and provide ample testing
time outside of standard operations. 
Do not operate with un-tested changes.  
Allow a mechanism for reverting software versions.  
Software engineers exist for a reason. Use them, but guide 
them with scientists and end-users.

\section{Written Procedures}
We quickly learned
that it was very important to have well-written, thorough operational 
procedures.
These served several purposes:
\begin{itemize}
\item Help ensure uniformity in procedures (hence data) across observers.
\item Serve as memory aids for rarely-done, or not recently performed tasks.
\item Help speed training of new observers.
\item Document existing procedures for evaluation by other project members in
light of their own areas of expertise.
\end{itemize}

These procedures continued to be developed and modified throughout the entire
survey.  In addition, we found
there were also a lot of little tidbits of information we called folklore
which didn't seem to fit in any
strictly-defined procedure.  We thus created a web-based system to allow
the easy addition and editing of these small bits of information. They were
broadly categorized into many different areas and like the standard
procedures, were completely searchable for easier retrieval.  Sometimes,
these bits of folklore became so numerous or complex that we could then
gather them together and form a formal procedure to replace the folklore
category.  This  mechanism turned out to be a good way to document
valuable information that seemingly had no other place for storage.

Lesson: Procedures and documentation written by the developers is a start,
but for real utility, they need to be continually modified and improved by
those actually using them.  A version-controlled documentation repository
provided this facility for us.  
Finding places for small bits of information that don't make a manual or
procedure by themselves can be quite important.

\section{Communication and Bug Reporting}
As with most human endeavors, communication in any astronomical project is
very important. More specifically, {\it open} and {\it abundant}
communication is very important.  Sometimes this idea runs contrary to
people's inclinations, but
like many in the business world have discovered (eg.
http://blog.redfin.com, Schwartz 2005),  we found
talking openly about our problems and weaknesses as well as our successes
and strengths
actually helped, not hurt, the project. Firstly, it built a sense of trust
and honesty among our participants and the outside world. And secondly,
it opened us up to insights and solutions from a much larger body of people than
might otherwise have been involved.  And finally, it kept the
collaboration involved and informed about the project.  

Open communications means  everyone should feel
free to discuss problems, solutions and other concerns. Hiding problems
behind private mailing lists, closed doors, and secret communiqu\'e only
increases the morale-zapping rumor mill and decreases the number of
possible solutions that can be offered.  The SDSS produced free, detailed,
public nightlogs from the observers, offered a plethora of 
archived mailing lists for all aspects of the project, held regular (often
weekly) phone meetings to discuss operations, controls, engineering, data
reduction and distribution, etc., and held bi-annual general
collaboration and engineering planning meetings which allowed for
significant face to face time amongst the participants.  Project management
routinely traveled from site to site in the collaboration, meeting people
in their home space, and learning the environment in which the team's
members operated.  We encouraged wide participation in both the bi-annual
meetings, including people involved in operations and data
reduction/quality in addition to the project engineers, scientists and
management.
Such attendance helped place the project, workload, and any coming
changes into context.

The nightlogs, mail archive, meetings, and an openly-accessible problem
reporting system all worked together to establish good communications
throughout the project.  Within the atmosphere this effort created, people
could freely discuss problems as things to be solved  not as things
requiring blame to be placed.  If a mistake was made, the system allowed
that mistake to be made and thus the system could be improved to help
prevent future similar mistakes.  Blame became less of an issue as positive
steps to solve problems became more of one.  

The mail archive and problem reporting system made finding past occurrences of
problems easier.  Providing a
project-level mail archive also meant each project member was relieved of the
responsibility of maintaining their own file system for the myriad of
project emails.  They could subscribe to the lists they wanted to, avoiding
the rest, thus keeping the flow of relevant information relatively high.   
The archive allowed us to locate old messages, search for
past events, and even scan critical lists for information without having to
wade through our own (usually overflowing) mail queue.

%The SDSS's detailed, public night logs were very important to the project.
%For one, they served a documentation purpose, recording the active
%observers, software versions, weather conditions etc., that would be needed
%later to understand or debug the obtained data.  Second, by establishing a
%culture of openness, the nightlogs unabashedly documented problems and
%mistakes, allowing the project later to generate solutions.\footnote{It is 
%amusing to note that mistakes mentioned in the observing
%nightlogs are usually traceable to the observer that made them: if the
%person that wrote the log used the word {\it we}, then the mistake was
%likely made by the other observer; if instead the report said {\it I},
%then it was likely the logging observer.}
%They also allowed the entire project, particularly the software and
%hardware engineers,  a little insight into operations,
%thereby gaining a better understanding of the impact of their decisions and
%their work on basic operations.  

Open communication also pertains to data rights.  As a community, we've
taken practically all approaches possible with respect to data rights and
each has their strengths and weaknesses. We tend, however, to lean towards
public data rights as soon as possible, for the following
reasons: 
\begin{itemize}
\item We are in the science business. Disseminating our results is part of
what we do. Science progresses more quickly when data are released sooner.
\item In today's large projects, community support is vital for continued
funding and operations.  Such support is easier to gather when 
giving something freely in return.
\item The temptation to keep data exclusive as a benefit to the project's
members is large, but there will always be an insider edge regardless of
the data rights themselves.  Insiders will understand the structure,
quality, and limitations of the data better than anyone else.  You
also do not want to restrict your insiders' ability to do science by
restricting their possible set of collaborators.
\item These new databases are large, complex affairs\footnote[1]{So we should
also add a note here to budget for this effort carefully. Just like
software development, data distribution will take significant time and
resources and is not easily handled by an informal effort.}.  Having more
public data access helps them develop more quickly.  The SDSS had
a period of exclusive data rights for the collaboration prior to each data
release.  Until late in the project, however, the data were not as easily
accessible during this proprietary period as they were in the public
releases. This scenario is simply a reflection of the complexity of data
releases. It is most efficient not to split your resources into developing
proprietary and public data release systems.  \end{itemize}

For bug/problem reporting, we used a modified version of the GNU
GNATS\footnote{http://www.gnu.org/software/gnats/}
system.  This system allowed us to set up a variety of broad problem categories,
each with individually-assigned notification lists and auto-notification of
status changes to the submitter (this feedback is important to help show
the submitter that something is happening).
Any member of the collaboration
could sign up for any category's notification list and all could see
whichever problem reports they desired.  We modified the stock GNU system
to include a {\it Scheduling} category where we could place a problem report in
various approved or un-approved states.  We found this modification allowed
us to continue to promote problem reporting by everyone, but gave us more
control on the priorities of addressing them. Just because someone filed a
problem report doesn't necessarily mean people should drop all they're
doing to address it.  During the active
development phases of the project, we held regular meetings to review and
prioritize problem reports and their solutions, usually with input from the
problem submitters.

Lesson: Fostering open, free communications invests everyone in the
project, provides for easier problem troubleshooting, and involves more
people in the solutions than might otherwise be.  Email along with phone and video
conferences are important components of this effort, but do not adequately
substitute for general face to face meetings.  Searchable, common
archives of communications serve many important objectives and become
valuable resources of information.  Establish a culture of problem solving,
not blame placing.  Build a sense of community and involvement. Evaluate your data rights policy carefully. Proprietary
rights are not as obviously beneficial as they may seem. Openly learn from
your mistakes.

\section{Stability vs. Change}
A survey or other long term project values stability.  Research science, however, values
continuous change and evolutionary improvements.  Balancing these two
motivations was one of the most difficult and time-consuming aspects of the
SDSS.  While we can't offer a short recipe for providing this balance,
we instead offer a few points for consideration.
\begin{itemize}
\item A full understanding of the scientific necessity and operational
impact of a proposed change is vital for evaluating its worth.
\item There are almost always unseen impacts from change.  Tread carefully
and thoroughly. The regression tests you already put in place will serve
you well here.
\item If a change is decided upon, don't be afraid to change back if things
don't go well. Corollary: make it easy to change back after initiating a
change.
\end{itemize}

Establishing a series of change control boards at different levels of the
project can help maintain both focus and a cross-discipline outlook when
evaluating change proposals. It also lengthens the timescale needed to
enact a change. At early levels of the project, this characteristic can be
a detriment, but later on, it can be a definite benefit.

\section{Conclusions}
The SDSS was by no means the first such successful large project in this
new era of astronomy, but it was arguably one with the most significant
impact relative to its cost \cite{tri05}. This success was due to several
simultaneous efforts, most of which are discussed here:
choosing and maintaining specific, clear, and interesting objectives 
(Sections 2 and 10), careful attention to project metrics and efficiencies
(Section 3 and 4), a large, cooperative collaboration of dedicated people
(Sections 5 and 9), and efficient hardware, software, and procedures
designed to work together from the start (Sections 1, 4, 6, 7, and 8).

Astronomy's new projects must maintain and inspire both their collaboration
participants and their funders.  They must attract top talent
and keep them with the project. They must avoid fatal miscalculations that
could cost long delays and increasingly larger amounts of money.  While the
SDSS did not invent the ideas presented here, it did at least end up
applying them reasonably well and we believe the key points will be relevant
to many future projects trying hard to ensure their early success.

We discussed many issues in this paper, but
the three areas we believe are most key are:
\begin{itemize}
\item {\bf Addressing the scientist dilemma:} You simply cannot hire
scientists into 100\% service positions and expect either to reap their full
benefit or keep them.  Scientists are not always the best managers
and the best managers are not always scientists.  Your project needs both.
There are various solutions to these problems, but they cannot be ignored.
\item {\bf Software:} We expect more and more out of our software which
in turn does more and more for us.  Even small efficiency increases made
through software that is repeatedly run can benefit the bottom line
immensely. The field of software engineering developed because the days of
a single person in front of a keyboard and screen turning out high quality
innovative software solutions in a few days are largely over.  Software needs
should be carefully evaluated from the beginning, including getting user
input, providing documentation and handover material, maintenance, and testing.
Plan ahead for continued software development after delivery.
\item {\bf Communication:} As in any human endeavor, excellent
communications are vital for astronomical projects.
A temptation is often
present to downplay and hide potentially negative news and to keep valuable
information private, both in an attempt to increase one's own value to the 
project and the project's value to the community. Both these
temptations must be overcome.  Your collaboration exists for a reason ---
trust them, use them, {\it include} them.  The better people are informed, the
more involved they become and the more they contribute.
%Your
%money-providing sponsors did not get their money because they were fooled by
%overly-superficial reports and rosy scenarios.   Honesty and
%forthrightness is nearly always repaid with increased support, understanding,
%and work.  
Establish a teamwork culture of solving problems together, not
one of assigning blame or assuring job security through information hoarding.
Evaluate your data distribution policy carefully.
\end{itemize}

The suggestions and solutions presented here may not work for every project
in every situation.  The issues
they address, however, are universal and any successful project will
have to address them to ensure its success.  We hope that if nothing
else, this work will help raise these important issues and inspire you
to find your own solutions that keep astronomy growing and flourishing
deep into this new era.

\acknowledgments
Funding for the SDSS and SDSS-II has been provided by the Alfred P. Sloan
Foundation, the Participating Institutions, the National Science
Foundation, the U.S. Department of Energy, the National Aeronautics and
Space Administration, the Japanese Monbukagakusho, the Max Planck Society,
and the Higher Education Funding Council for England. The SDSS Web Site is
http://www.sdss.org/.

The SDSS is managed by the Astrophysical Research Consortium for the
Participating Institutions. The Participating Institutions are the American
Museum of Natural History, Astrophysical Institute Potsdam, University of
Basel, University of Cambridge, Case Western Reserve University, University
of Chicago, Drexel University, Fermilab, the Institute for Advanced Study,
the Japan Participation Group, Johns Hopkins University, the Joint
Institute for Nuclear Astrophysics, the Kavli Institute for Particle
Astrophysics and Cosmology, the Korean Scientist Group, the Chinese Academy
of Sciences (LAMOST), Los Alamos National Laboratory, the
Max-Planck-Institute for Astronomy (MPIA), the Max-Planck-Institute for
Astrophysics (MPA), New Mexico State University, Ohio State University,
University of Pittsburgh, University of Portsmouth, Princeton University,
the United States Naval Observatory, and the University of Washington. 

%use bib.bib for bibliography
\bibliography{bib}
\bibliographystyle{spiebib}

\end{document}